\documentclass[twocolumn,amsmath,showpacs,amssymb,aps,pra,floatfix,nofootinbib]{revtex4}
\usepackage{graphicx,psfrag}
\usepackage{graphics}
\usepackage{bm}
\begin{document}
\title{On the Phase Covariant Quantum Cloning}
\author{V. Karimipour } \email{vahid@sharif.edu}
\author{A. T. Rezakhani } \email{tayefehr@mehr.sharif.edu}
\affiliation{Department of Physics, Sharif University of
Technology, P.O. Box 11365-9161, Tehran, Iran}
\begin{abstract}
It is known that in phase covariant quantum cloning the equatorial
states on the Bloch sphere can be cloned with a fidelity higher
than the optimal bound established for universal quantum cloning.
We generalize this concept to include other states on the Bloch
sphere with a definite $z$ component of spin. It is shown that
once we know the $z$ component, we can always clone a state with a
fidelity higher than the universal value and that of equatorial
states. We also make a detailed study of the entanglement
properties of the output copies and show that the equatorial
states are the only states which give rise to separable density
matrix for the outputs.
\end{abstract}
 \maketitle

\section{Introduction}
Universal quantum cloning refers to the possibility of
constructing unitary transformations which approximately copy an
arbitrary quantum state and hence partially alleviate the
limitations of the no-cloning theorem \cite{wz} (see also \cite{p}
and \cite{gw}). It was first achieved by Bu\v{z}ek and Hillery in
\cite{bh1} in which they proposed a cloning transformation which
clones arbitrary states with equal fidelity $\frac{5}{6} \simeq
0.83 $. Their pioneering work stimulated a lot of intense research
in quantum cloning, a sample of which includes works on proofs of
optimality \cite{bem,bdefms,Dariano,Fiurazek}, generalizations to
$N\rightarrow M$ cloning \cite{gm}, cloning of $d$-level states
\cite{w,z,bh2,fmw}, and finally experimental realization of
cloning by various techniques
\cite{cjfsmpj,lshb,fgrsz}.\\

Since the optimal bound of $5/6$ for fidelity was set for
universal cloning, attempts were also made to go beyond this limit
by cloning special subsets of states for which we have some a
priori partial information. This search was indeed successful and
led to the so-called phase covariant quantum cloning
\cite{bcdm,fmww,fimw,Dariano,cdg}. For two level states, or
qubits, phase covariant quantum cloning means that a certain class
of states, called ($x-y$) equatorial states, defined as
\begin{eqnarray}\label{phasecovariant}
&|\psi\rangle = \frac{1}{\sqrt{2}}(|0\rangle + e^{i\phi}
|1\rangle)
\end{eqnarray}
can be cloned with a fidelity $ F= \frac{1}{2}(1+
\frac{1}{\sqrt{2}}) \simeq 0.85 $ which is slightly higher than
optimal bound achievable for universal quantum cloning. For
$d$-level states it means that states of the form
\begin{eqnarray}\label{phasecovariantdlevel}
&|\psi\rangle = \frac{1}{\sqrt{d}}\sum_{k=0}^{d-1} e^{i\phi_k}
|k\rangle
\end{eqnarray}
can be cloned with a fidelity \cite{fimw}
\begin{eqnarray}
&F=\frac{1}{d}+\frac{1}{4d}(d-2+\sqrt{d^2+4d-4})>F_{\rm{univ.}}=\frac{d+3}{2(d+1)}.\hskip
2mm
\end{eqnarray}
The crucial property of this class which allows for this higher
fidelity is that all the coefficients in their expansion have
equal norm. Due to this property a state dependent term in the
final density matrix of the clones in the cloning transformation,
which is of the form $\sum_{k} |\alpha_k|^2 |k\rangle \langle k|
$, becomes automatically state independent (universal), hence no
need for making its coefficient vanish by tuning the parameters of
the cloning transformation. With the automatic disappearance of
this term and one more parameter at hand we find the chance to
obtain higher fidelity than the optimal one. This is all the
technical point of the phase covariant quantum cloning. There is
of course one motivation for studying these states which comes
from quantum cryptography, since at least in the BB84 protocol,
the states in transfer between the legitimate parties are of this
form and an eavesdropper needs only to clone
these kinds of states to threat the security of the communication.\\

However, when we think in terms of physical properties, the
partial information that we have about these states is that the
$z$ component of their spin is zero. Therefore, it is natural to
ask a more general question, that is, how well we can clone a spin
states $ |\psi \rangle $ if we know the third component of its
spin $ \langle \psi|\sigma_z| \psi \rangle $? This question is
specially interesting for those who try to achieve optimal cloning
by NMR techniques \cite{cjfsmpj}. In fact this is precisely the
state of a nuclear spin which is precessing in magnetic field with
a definite energy. In this sense we not only generalize the
concept of phase covariant quantum cloning, but describe it in a
physically and experimentally interesting context.

We show that there exist a one-parameter family of cloning
transformations in which by tuning the parameter one can always
clone such states with higher fidelity than the optimal one.
Furthermore we show that within this class, the case of equatorial
states give a lower fidelity of cloning compared to other states.
However they are unique in the sense that they are the only states
in this class which give rise to separable density matrix for the
outputs copies. We also show that our consideration can be readily
generalized to $d$-level states.\\
The structure of this paper is as follows: In Sec. {\ref{sec2}} we
study the general properties of a one parameter family of cloning
transformations of qubits. In Sec. {\ref{sec3}} we make detailed
comparison between different cloning transformations, namely the
universal cloning machine proposed by Bu\v{z}ek and Hillery, the
phase covariant cloning proposed in \cite{bcdm} and the one
proposed in this paper. In Sec. {\ref{sec4}} we briefly discuss
the phase covariant cloning of $d-$ level states \cite{fimw} in
this new context. The paper ends with a conclusion.
\section{Cloning transformation of qubits: general
properties}\label{sec2} Consider the following cloning
transformation
\begin{eqnarray}\label{basiccloner}
&&U: |0\rangle_{a}\longrightarrow \nu |0,0\rangle_{a,b}
|0\rangle_{x} + \mu
  (|01\rangle+ |10\rangle)_{a,b}
  |1\rangle_{x} \cr
  &&U: |1\rangle_{a} \longrightarrow \nu |1,1\rangle_{a,b} |1\rangle_{x} + \mu
  (|01\rangle+ |10\rangle)_{a,b}
  |0\rangle_{x}
\end{eqnarray}
where on the left hand side we have not shown the blank state and
initial state of the cloning machine and on the right hand side,
the states from left to right correspond respectively to the input
($a$), the copy ($b$) and the machine states ($x$). The states
$|0\rangle $ and $|1\rangle$, are also orthonormal regardless of
their indices. The only requirement for this transformation to be
unitary is that $ \mu $ and $ \nu $ be related as
\begin{eqnarray}\label{unitarity}
&\nu^2 + 2 \mu^2 = 1.
\end{eqnarray}
Consider now a general two level state, i.e. a state with a
definite spin in the direction ${\bf n}=(\sin \theta \cos \phi,
\sin \theta \sin \phi, \cos \theta)$, where $ \theta $ and $\phi$
are the polar coordinates on the unit sphere. This state has the
following form in the $z$-basis ($ \sigma_{z} |0\rangle =
|0\rangle,  \sigma_{z} |1\rangle = - |1\rangle $)

\begin{eqnarray}\label{state}
&|{\bf n}\rangle = \cos \frac{\theta}{2}|0\rangle +  e^{i\phi}\sin
\frac{\theta}{2}|1\rangle
\end{eqnarray}

The output state of the composite system $ab$ is obtained by
tracing out the states of the machine $x$, that is

\begin{eqnarray}\label{outputrho}
  &\rho_{ab}^{(out)} = {\text{Tr}}_{x} (U|{\bf n}\rangle \langle {\bf
  n}|U^{\dagger}).
\end{eqnarray}
When acted on by the cloning machine (\ref{basiccloner}) this
state gives rise to the following density matrix for the output
$a$
\begin{eqnarray}\label{rhoaoutbad}
\rho_{a}^{(out)} = &\mu^2& 1 + 2\mu \nu |\psi\rangle \langle \psi|
\nonumber\\&+& (\nu^2-2\mu\nu)(\cos^2
\frac{\theta}{2}|0\rangle\langle 0|+ \sin^2
\frac{\theta}{2}|1\rangle\langle 1|).\hskip 2mm
\end{eqnarray}
The new copy $ b$ will also have the same density matrix. The
fidelity of cloning defined by $ F:=\langle {\bf n}|
\rho_a^{(out)}|{\bf n}\rangle $ is found to be
\begin{eqnarray}\label{basicfidelity1}
 &F (\theta)= \mu^2 + 2\mu \nu  + (\nu^2-2\mu\nu) (\cos^4 \frac{\theta}{2} + \sin^4 \frac{\theta}{2} )
\end{eqnarray}
which after a little algebra using, the fact that $ \cos \theta =
\langle {\bf n} |\sigma_z|{\bf n} \rangle\equiv \langle \sigma_z
\rangle $, and the normalization condition $ \nu^2 + 2\mu^2 = 1 $,
takes the form
\begin{eqnarray}\label{basicfidelity2}
 F (\theta) &=& \frac{1}{2} + \mu\nu + ( \frac{\nu^2}{2}-\mu \nu) \cos^2
 \theta\nonumber\\
 &=& \frac{1}{2} + \mu\nu + ( \frac{\nu^2}{2}-\mu \nu)  \langle \sigma_z \rangle^2.
 \end{eqnarray}

The last term clearly depends on the input state. All the states
on the Bloch sphere with the same value of $ \phi $ are cloned
with equal fidelity, a special subclass of these states are the so
called equatorial states, those with $ \langle \sigma_z \rangle =
0 $.

Following Bu\v{z}ek and Hillery {\cite{bh1}} it is useful to
define and calculate two distances which characterize further the
quality of cloning,
 namely
\begin{eqnarray}\label{Dab1}
  &D^{(1)}_{ab}:= {\text{Tr}}\Big[(\rho_{ab}^{(out)} - \rho_a^{(out)}\otimes
  \rho_b^{(out)})^2\Big]
\end{eqnarray}
which measures the degree of entanglement of the two output copies
and
\begin{eqnarray}\label{Dab2}
  &D^{(2)}_{ab}:= {\text{Tr}}\Big[(\rho_{ab}^{(out)} - \rho_a^{(id)}\otimes
  \rho_b^{(id)})^2\Big],
\end{eqnarray}
which measures the distance of the two mode output density matrix
with the ideal situation of having two disentangled exact copies
of the input states.

The calculation of these distances are straightforward but rather
lengthy. We give only the final results
\begin{eqnarray}\label{Dab2final}
D^{(1)}_{ab}(\theta)=&A_8(\mu) \cos^8 \frac{\theta}{2}+A_6(\mu)
\cos^6 \frac{\theta}{2}\nonumber\\&+A_4(\mu) \cos^4
\frac{\theta}{2}+A_2(\mu) \cos^2 \frac{\theta}{2}+A_0(\mu)
\end{eqnarray} where

\begin{eqnarray}\label{amus}
&&A_8(\mu)= 576 \mu^8 - 768 \mu^6 + 352 \mu^4 - 64 \mu^2 + 4 \cr
&&A_6(\mu)= -1152 \mu^8 + 1536 \mu^6 -704 \mu^4 + 128 \mu^2 -8\cr
&&A_4(\mu)= 672 \mu^8 - 928 \mu^6 + 424 \mu^4 - 72 \mu^2 + 4\cr
&&A_2(\mu)= -96 \mu^8 + 160 \mu^6 - 72 \mu^4 + 8 \mu^2\cr
&&A_0(\mu)= 4\mu^8 + 2 \mu^4
\end{eqnarray}
and

\begin{eqnarray}\label{Dab1final}
&D^{(2)}_{ab}(\theta)= 8 \mu^4 - (6 \mu^4 +  \mu^2+ 2 \mu \nu -
1)\sin^2
  \theta
\end{eqnarray}

\section{Comparison of cloning machines}\label{sec3}
Until now the value of $ \mu $ has been kept arbitrary. We should
now fix it and hence complete definition of our cloning
transformation (\ref{basiccloner}). In the sequel, we consider
three different cases.
\subsection{Universal quantum cloning}
Looking at Eq. (\ref{basicfidelity2}), we find that universality,
in the sense of Bu\v{z}ek and Hillery, is achieved only be setting
$ \frac{\nu^2}{2}-\mu \nu = 0 $ which together with normalization
yields

\begin{eqnarray}\label{bhclone}
&\mu = \frac{1}{\sqrt{6}}\hskip 1.5cm
F_{\rm{optimal}}^{\rm{universal}} =
  \frac{5}{6} \simeq  0.833.
\end{eqnarray}
Here no optimization should be performed, since the demand of
universality has fixed completely the parameter $ \mu$. It is
interesting to note that in this case the two distances $
D_{ab}^{(1)} $ and $ D_{ab}^{(2)} $ are also state independent. In
fact, by inserting the above value for $ \mu $ in Eqs.
(\ref{Dab1final}) and (\ref{Dab2final}) one finds that

\begin{eqnarray}\label{dabBH}
&D_{ab}^{(1)}(\theta) = \frac{19}{324} \hskip 1cm
D_{ab}^{(2)}(\theta) = \frac{2}{9}
  \hskip .5cm \forall \ \ \theta.
\end{eqnarray}
\subsection{Phase covariant quantum cloning}\label{sec3b}
In this part we are interested in cloning only the states with $
\langle \sigma_z \rangle = \cos \theta = 0 $. Thus the parameter $
\mu $ is free and we can fix it by maximizing the value of $
F(\frac{\pi}{2}) = \frac{1}{2} + \mu \nu = \frac{1}{2} + \mu
\sqrt{1-2 \mu^2}$. One thus finds
\begin{eqnarray}\label{fidelityPC}
\mu = \frac{1}{2}\hskip 1.2cm
 &F_{\rm{opt.}}^{\frac{\pi}{2}} =
 \frac{1}{2}(1 + \frac{1}{\sqrt{2}}) \simeq 0.854
\end{eqnarray}
which is slightly higher than the value for universal quantum
cloning.\\
The distances are found to be
\begin{eqnarray}\label{dabPC}
&D_{ab}^{(1)}(\frac{\pi}{2}) = \frac{9}{64}    \hskip 1cm
D_{ab}^{(2)}(\frac{\pi}{2})= \frac{7}{8}-\frac{1}{\sqrt {2}}.
\end{eqnarray}
Although the distance $D_{ab}^{(1)}$ for the equatorial states is
appreciably higher than the universal value, as we will see below
the equatorial states are separable when cloned phase covariantly
\cite{fmww} while in universal cloning machine of Bu\v{z}ek and
Hillery the output states are not separable.

\subsection{Cloning of states with a definite
component of spin along the $z$ direction}\label{sec3c}

\begin{figure}[ht]
\psfrag{f1}[Bc][][0.75][0]{$F_{opt.}(\theta)$}
\psfrag{g1}[Bc][][0.75][0]{$F_{univ.}$}
\psfrag{t1}[Bc][][0.75][0]{$\theta{\text{(rad)}}$}
\psfrag{h1}[Bc][][0.75][0]{$F_{opt.}$}
\includegraphics[width=6cm,height=5cm]{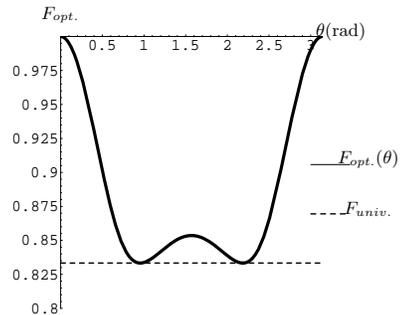}
\caption{The optimal fidelity of cloning a qubit with known $z$
component as a function of $\theta.$} \label{fig1}
\end{figure}

\begin{figure}[ht]
\psfrag{f2}[Bc][][0.75][0]{$D^{(1)}(\theta)$}
\psfrag{g2}[Bc][][0.75][0]{$D^{(1)}_{univ.}$}
\psfrag{t2}[Bc][][0.75][0]{$\theta{\text{(rad)}}$}
\psfrag{h2}[Bc][][0.75][0]{$D^{(1)}$}
\includegraphics[width=7cm,height=5.3cm]{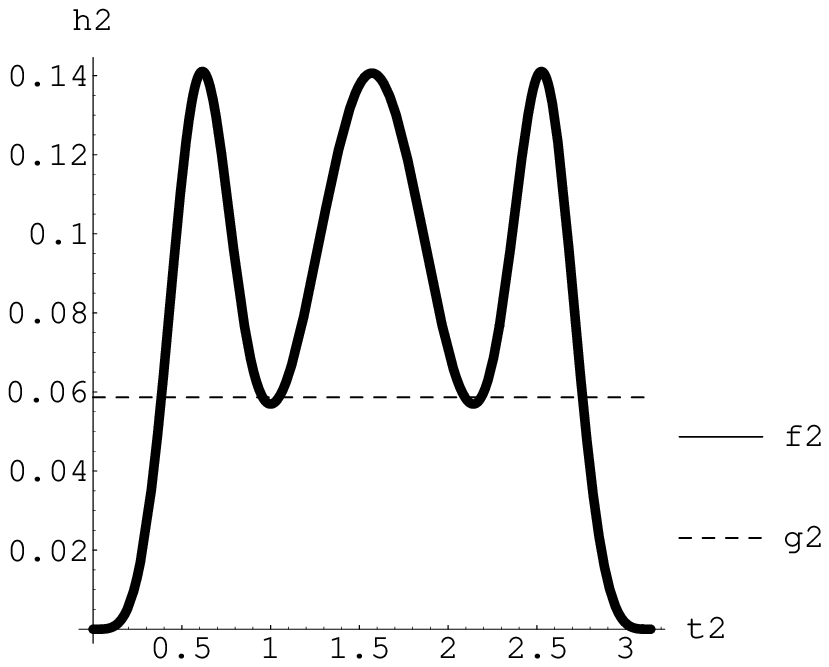}
\caption{The distance $D_{ab}^{(1)}$ as a function of $\theta$. }
\label{fig2}
\end{figure}

\begin{figure}[ht]
\psfrag{f3}[Bc][][0.75][0]{$D^{(2)}(\theta)$}
\psfrag{g3}[Bc][][0.75][0]{$D^{(2)}_{univ.}$}
\psfrag{t3}[Bc][][0.75][0]{$\theta{\text{(rad)}}$}
\psfrag{h3}[Bc][][0.75][0]{$D^{(2)}$}
\includegraphics[width=7cm,height=5.3cm]{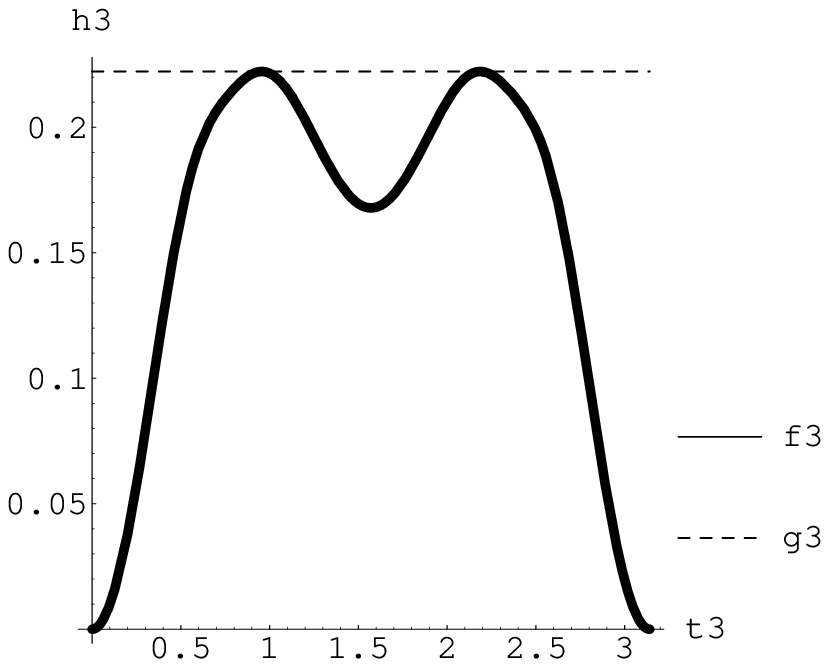}
\caption{The distance $D_{ab}^{(2)}$ as a function of $\theta$. }
\label{fig3}
\end{figure}

In this case, we fix the value of $ \theta $ and find from Eq.
(\ref{basicfidelity2}) that $F$ is extremized by two values of $
\mu $ obtained from

\begin{eqnarray}\label{muoptimal}
&\tan^2 \theta = \frac{2\mu \sqrt{1-2\mu^2}}{1-4\mu^2}\hskip .3cm
  {\rm {or}} \hskip .3cm \mu^2 = \frac{1}{4}\big( 1\pm \frac{1}{\sqrt{1+2\tan^4
  \theta}}\big).
\end{eqnarray}
It turns out that the negative sign corresponds to the maximum
fidelity. Inserting this value of $ \mu$ in Eqs.
(\ref{basicfidelity2}), (\ref{Dab1final}) and (\ref{Dab2final})
will give us the optimal fidelity and the distances for this class
of states. The results are shown in Figs. {\ref{fig1}},
{\ref{fig2}} and {\ref{fig3}}. It is seen clearly that for each
fixed $ \theta $, one can clone the spin state with a fidelity
greater than the universal value and for most angles a higher
fidelity can be obtained than for the equatorial states. If judged
on the basis of the distances $ D_{ab}^{(1)} $ and $D_{ab}^{(2)}$,
it also appears from Figs. {\ref{fig2}} and {\ref{fig3}}, that
there are other states which are closer to a product state than
the equatorial ones. However the equatorial states are unique in
one important respect which is discussed in the next subsection.
\vskip .1cm
\vskip -.9cm
\vskip -.9cm
\subsection{Separability properties}\label{sec3d}
\begin{figure}[ht]
\psfrag{h4}[Bc][Bc][0.9][0]{\it{nonpositive eigenvalue}}
\psfrag{t4}[Bc][][0.9][0]{$\theta{\text{(rad)}}$}
\includegraphics[width=7cm,height=5.3cm]{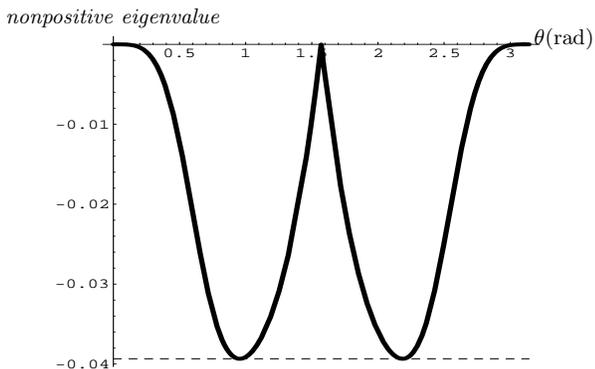}
\caption{The nonpositive eigenvalue of $[\rho_{ab}^{(out)}]^{T_a}$
vs. $\theta$. The dashed line shows its value in universal
cloning.} \label{fig4}
\end{figure}

For the universal cloning case, using Peres-Horodecki's positive
partial transposition (PPT) criterion {\cite{peres,horodecki}}, it
has been shown that two output modes are inseparable, while the
phase covariant cloning of equatorial states lead to separable
copies \cite{fmww}. To check separability for general angles, we
have numerically computed the eigenvalues of the partial transpose
of the output density matrix (\ref{outputrho}) which is of the
following form
\begin{widetext}
\begin{eqnarray}\label{outputrhoexplicit}
&[\rho_{ab}^{(out)}]^{T_a} = \left(
\begin{array}{cccc}
 \nu^2 \cos^{2}\frac{\theta}{2} & \mu \nu e^{-i\phi}\sin \frac{\theta}{2}\cos \frac{\theta}{2} & \mu \nu e^{i\phi}\sin \frac{\theta}{2}\cos \frac{\theta}{2} & \mu^2\\
  \mu \nu e^{i\phi}\sin \frac{\theta}{2}\cos \frac{\theta}{2} & \mu^2 & 0 &\mu \nu e^{i\phi}\sin \frac{\theta}{2}\cos \frac{\theta}{2} \\
  \mu \nu e^{-i\phi}\sin \frac{\theta}{2}\cos \frac{\theta}{2} & 0 & \mu^2 & \mu \nu e^{-i\phi}\sin \frac{\theta}{2}\cos \frac{\theta}{2} \\
  \mu^2 & \mu \nu e^{-i\phi}\sin \frac{\theta}{2}\cos \frac{\theta}{2} & \mu \nu e^{i\phi}\sin \frac{\theta}{2}\cos \frac{\theta}{2} & \nu^2 \sin^2 \frac{\theta}{2}
\end{array}\right),
\end{eqnarray}
\end{widetext}
and have found that three of the eigenvalues are always positive,
while one of them is marginally negative and becomes zero only for
the states on the north pole $|0\rangle $, the south pole
$|1\rangle $ and the equator of the Bloch sphere. The values of
this negative eigenvalue is shown in Fig. {\ref{fig4}} and is
compared to the negative eigenvalue of the universal machine which
is $\frac{1}{3}-\frac{\sqrt{5}}{6} \simeq - 0.04 $, (while the
other eigenvalues of the universal cloning machine being
$\frac{1}{6}, \frac{1}{6} $ and $\frac{1}{3}+\frac{\sqrt{5}}{6}$,
all independent of $ \theta $ ).

Thus also in this general class of states, the equatorial states
are special in that they are completely separable. However, if one
considers the multiple criteria of high fidelity and approximate
separability then it may be concluded from all the above figures
that the states with angles less than $ \theta \leq  0.5 $ radians
around the north and south poles can be cloned with sufficiently
high (larger than 0.9 fidelity) and rather good separability
properties.
\subsection{Optimality}
In this section we address the question of optimality of the
transformations (\ref{basiccloner}). The general form of these
transformations are the same as the original cloning
transformations found by Bu\v{z}ek and Hillery \cite{bh1} and
proved to be optimal for universal cloning
\cite{bem,bdefms,Dariano,Fiurazek}. Here we have shown that by
adjusting the single parameter of these transformations, one can
clone states with definite $z$ components of spin, with a higher
than universal fidelity. However it may be possible to go beyond
these one parameter family of transformations and obtain even
higher fidelity. There is in fact a constructive procedure for
deriving the trace preserving completely positive (CP) maps which
perform a given task \cite{Fiurazek} like cloning, to the best
approximation. However we think that by following the procedure of
\cite{Fiurazek} our transformations may not retain their simple
form that they have now.

\section{Generalization to d-level states}\label{sec4}
Since phase covariant quantum cloning has been achieved for
$d$-level states, again with a fidelity which is higher than the
universal value, it is natural to ask how the above considerations
extend to $d$-level states. Consider the following cloning
transformation \cite{fimw}, which is a simple and natural
generalization of (\ref{basiccloner})
\begin{eqnarray}\label{basicclonerd}
&|j\rangle \rightarrow \nu |j,j\rangle_{a,b} |j\rangle_x +\mu
\sum_{l\neq
  j}(|j,l\rangle + |l,j\rangle)_{a,b}|l\rangle_x.
\end{eqnarray}
It can be easily verified that this transformation is unitary
provided that we have
\begin{eqnarray}\label{norm2}
&\nu^2 + 2(d-1)\mu^2 = 1.
\end{eqnarray}
In particular for 3-level states or qutrits, the transformation is
\begin{eqnarray}\label{basiccloner3}
|0\rangle &\rightarrow & \nu |0,0\rangle |0\rangle +  \mu
(|0,1\rangle + |1,0\rangle)|1\rangle +
  \mu (|0,2\rangle + |2,0\rangle)|2\rangle \nonumber\\
  |1\rangle &\rightarrow & \nu |1,1\rangle |1\rangle +  \mu (|0,1\rangle + |1,0\rangle)|0\rangle +
  \mu (|1,2\rangle + |2,1\rangle)|2\rangle \nonumber\\
  |2\rangle &\rightarrow & \nu |2,2\rangle |2\rangle +  \mu (|0,2\rangle + |2,0\rangle)|0\rangle +
  \mu (|1,2\rangle +  |2,1\rangle)|1\rangle\nonumber\\
\end{eqnarray}

The cloning transformation (\ref{basicclonerd}) transforms a pure
state
\begin{eqnarray}\label{stated}
&|\psi\rangle = \sum_{j=0}^{d-1} \alpha_j |j\rangle
\end{eqnarray}
into a mixed state
\begin{eqnarray}\label{outputd}
&\rho_{a}^{(out)} = \mu^2 1 + ((d-2)\mu^2 + 2 \mu
\nu)|\psi\rangle\langle \psi|\nonumber\\ &+ (\nu^2 - 2\mu \nu)
\sum_{l} |\alpha_l|^2 |l\rangle \langle l|
\end{eqnarray}
and in phase covariant cloning, one gets rid of the final term by
considering only the equatorial states of the form given in Eq.
(\ref{phasecovariantdlevel}). Clearly this is a heavy restriction
on the states. To see what this implies for the states in terms of
observables we note that the Lie algebra of $SU(d)$ is spanned by
traceless hermitian matrices. The Cartan subalgebra of this Lie
algebra which generalizes the $ \sigma_z $ Pauli matrix for spins,
is spanned by diagonal traceless matrices, $H_1, H_2, \cdots
H_{d-1}$ normalized to ${\text{tr}} H_i H_j = \delta_{ij}$. One
convenient choice is $ H_{k} = \frac{1}{\sqrt{k(k+1)}}{\rm
{diagonal}} (1, 1, \cdots, 1, -k, 0, 0, \cdots 0)$. For example,
for qutrits we have
\begin{eqnarray}\label{cartan}
  &H_1 = \frac{1}{\sqrt{2}}\left(\begin{array}{ccc}
    1 &  &  \\
     & -1 &  \\
     & & 0 \
  \end{array}
  \right) \hskip .2cm {\rm and } \hskip .2cm  H_2 = \frac{1}{\sqrt{6}}\left(\begin{array}{ccc}
    1 &  &  \\
     & 1 &  \\
     & & -2 \
  \end{array}
  \right).\hskip 2mm
\end{eqnarray}
Thus phase covariant qutrit states are precisely those states for
which $ \langle \psi| H_1 |\psi \rangle = \langle \psi| H_2 |\psi
\rangle = 0. $ In fact the most general state of a qutrit is given
by
\begin{eqnarray}\label{qutrit}
  &|\psi\rangle = \cos \theta |0\rangle + \sin\theta \cos\phi
  e^{i\alpha} |1\rangle + \sin\theta \sin\phi
  e^{i\beta} |1\rangle \hskip 2mm
\end{eqnarray}
and the fidelity of cloning of this state by the transformation
(\ref{basiccloner3}) is found from Eq. (\ref{outputd}) to be equal
to
\begin{eqnarray}\label{fidelity3}
&F= \langle \psi| \rho_{a}^{(out)}|\psi\rangle = 2 \mu^2 + 2
  \mu \nu + (\nu^2 - 2 \mu \nu) A_{\psi}\hskip 2mm
\end{eqnarray}
where
\begin{eqnarray}
&A_{\psi}:= \sum_{k=0}^{2} |\alpha_k|^4 = \cos^4\theta + \sin^4
\theta(\cos^4 \phi + \sin^4 \phi)\hskip 3mm
\end{eqnarray}

\begin{figure}[Ht]
\psfrag{h5}[Bc][][0.75][0]{$F_{opt.}$}
\psfrag{t5}[Bc][][0.75][0]{$A_{\psi}$}
\psfrag{positive}[Bc][][0.75][0]{$positive$}
\psfrag{negative}[Bc][][0.75][0]{$negative$}
\includegraphics[width=7cm,height=5.3cm]{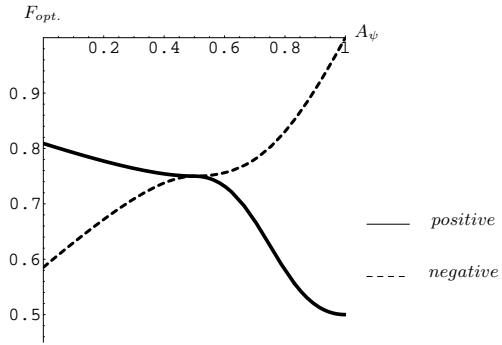}
\caption{The optimum fidelity of cloning a qutrit as a function of
$A$. The two curves correspond to different choices of the signs
for optimal $\mu$ (Eq. (\ref{mu3})).} \label{fig5}
\end{figure}

For a phase covariant state where all the coefficients have equal
amplitude we have $ \cos\theta = \frac{1}{\sqrt{3}}\ \ $  and  $\
\ \cos \phi = \sin \phi = \frac{1}{\sqrt 2}\ $. For this very
specific class of states with only two free parameters $ \alpha $
and $ \beta $, the fidelity is found from Eq. (\ref{fidelity3}) to
be $ \  F = 2 \mu^2+\frac{1}{3}\nu^2 + \frac{4}{3}\mu \nu \  $
which is optimized by taking $\  \mu^2 =
\frac{1}{8}(1+\frac{1}{\sqrt{17}})\ $ giving a value of $\
F_{\rm{optimal}} = \frac{1}{12}(5+\sqrt{17})\simeq 0.76 \ $
\cite{Dariano}. (There exists also another solution, namely $
\mu^2 = \frac{1}{8}(1-\frac{1}{\sqrt{17}})$, but for this
particular situation it is the first solution which gives the
higher fidelity, however see below and Fig. {\ref{fig5}}.) As
noted above these states are those for which $\langle H_1 \rangle
= \langle H_2 \rangle = 0 $.\\ However instead of restricting
oneself to this very specific class of states we can fix $
A_{\psi} $ and then optimize the fidelity. In this case one finds
that the optimum values of $ \mu ^2 $ are obtained from

\begin{eqnarray}\label{mu3}
&\mu^2 = \frac{1}{8}(1\pm \sqrt{\frac{\eta}{\eta+4}}),
\end{eqnarray}
where $ \eta \equiv\frac{(1-2A_{\psi})^2}{(1-A_{\psi})^2}$.
Reinserting these optimal values of $ \mu^2$ in Eq.
(\ref{fidelity3}) we obtain the optimal values of fidelity for
each value of $A_{\psi}$. The results are shown in Fig.
{\ref{fig5}}, where the two curves correspond to the two choices
of sign in the expression for $ \mu^2$. It is seen that for
obtaining the best possible cloning one should use either the plus
or the minus sign depending on the value of $A_{\psi}$. Finally we
observe that in general and for $d$ level states, the quantity $
A_{\psi}:= \sum_{k=0}^{d-1} |\alpha_k|^4 $, can actually be
expressed in terms of the expectation values  of the operators $
H_k$ in the form
\begin{eqnarray}\label{Ah}
&A_{\psi} = \frac{1}{d} + \sum_{k=1}^{d-1}\langle H_k
\rangle_{\psi}^2.
\end{eqnarray}

This equation shows that the equatorial states are a very
restricted class of states for which the expectation values of all
the observables $H_1 $ and $H_{d-1}$ have been fixed to zero. By
fixing the value of the quantity $ A_{\psi}$ which has the above
simple expression in terms of these observables one can clone a
much larger class of states with higher than universal fidelity.
In particular one sees that while for two level states there is
no difference in the number of parameters of the equatorial
states and states with non-zero $\langle \sigma_z\rangle$, the
difference in the number of free parameters in the general $d$
level case can be quite large depending on the value of $ d$.


\section{Discussion}
We have described the true physical context for phase covariant
quantum cloning, that is we have shown that once we have partial
information about a state like the $z$ component of spin or the
energy of a nuclear spin in a magnetic field, we can clone such a
state with a fidelity higher than the optimal universal fidelity
and higher than equatorial states. We have provided a one
parameter family of cloning transformation so that for each value
of the $z$ component, we can tune the parameter to obtain the
maximum fidelity.  We have also shown in this class the equatorial
states are the only ones which give rise to separable density
matrix for the outputs. However we have shown that it is possible
to clone all the states in the vicinity of the north and south
pole, for approximately ( $\theta < 0.5 $  radians or $ \pi -
\theta < 0.5 $ radians), with sufficiently high (larger than 0.9)
fidelity and rather good separability properties. The results of
this paper may be useful for those who are interested in
experimental realization of quantum cloning by using Nuclear
Magnetic Resonance (NMR) techniques. We have also discussed how
phase covariant quantum cloning of $d$-level states can be
generalized in the same way.

\begin{thebibliography}{99}


\bibitem{wz} W.K. Wootters, and W.H. Zurek, Nature (London) {\bf
299}, 802 (1982).

\bibitem{p} A. Peres, How the no cloning theorem got its name,
quant-ph/0205076.

\bibitem{gw} G.C. Ghirardi and T. Weber, Nuovo Cimento B {\bf
78}, 9 (1983). (The impossiblity of exact cloning has been first
mentioned in a referee report by G.C. Ghirardi (1981) in response
to a paper submitted to Foundations of Physics.)

\bibitem{bh1} V. Bu\v{z}ek and M. Hillery, Phys. Rev. A {\bf 54},
1844 (1996).

\bibitem{bem} D. Bru\ss, A.E. Ekert and C. Macchiavello, Phys.
Rev. Lett. {\bf 81}, 2598 (1998).

\bibitem{bdefms} D. Bru\ss, D. DiVincenzo, A.E. Ekert, C.A.
Fuchs, C. Macchaivello and J.A. Smolin, Phys. Rev. A {\bf 57},
2368 (1998).
\bibitem{Dariano} G.M. D'Ariano and P. Lo Presti, Phys. Rev. A {\bf 64}, 042308 (2001).
\bibitem{Fiurazek}
J.Fiurasek, Phys. Rev. A {\bf{64}}, 062310 (2001).
\bibitem{gm} N. Gissin and S. Massar, Phys. Rev. Lett. {\bf 79}, 2153 (1997).

\bibitem{w} R.F. Werner, Phys. Rev. A {\bf 58}, 1827 (1998); M.
Keyl and R.F. Werner, J. Math. Phys. {\bf 40}, 3283 (1999).

\bibitem{z} P. Zanardi, Phys. Rev. A {\bf 58}, 3448 (1998).

\bibitem{bh2} V. Bu\v{z}ek and M. Hillery, Phys. Rev. Lett. {\bf 81}, 5003 (1998).

\bibitem{fmw} H. Fan, K. Matsumoto and M. Wadati, Phys. Rev. A {\bf
64}, 064301 (2001).

\bibitem{cjfsmpj} H.K.
Cummins, C. Jones, A. Furze, N.F. Soffe, M. Mosca, J.M. Peach, and
J.A. Jones,  Approximate Quantum Cloning with Nuclear Magnetic
Resonance, quant-ph/0111098.

\bibitem{lshb} A. Lamas-Linares, C. Simon, J.C. Howell and D.
Bouwmeester, Science {\bf 296}, 712 (2002).

\bibitem{fgrsz}S. Fasel, N. Gisin, G. Ribordy, V. Scarani,
and H. Zbinden , Quantum cloning with an optical fiber amplifier,
quant-ph/0203056.

\bibitem{bcdm} D. Bru\ss, M. Cinchetti, G.M. D'Ariano and C.
Macchiavello, Phys. Rev. A {\bf 62}, 012302 (2000).

\bibitem{fmww} H. Fan, K. Matsumoto, X-B. Wang, and W. Wadati,
Phys. Rev. A {\bf 62}, 012304 (2002).

\bibitem{fimw} H. Fan, H. Imai, K. Matsumoto, and X-B.
Wang, Phase-covariant quantum cloning of qudits, quant-ph/0205126.



\bibitem{cdg} N.J. Cerf, T. Durt,  and N. Gisin
Cloning a Qutrit, quant-ph/0110092.

\bibitem{peres} A. Peres, Phys. Rev. Lett. {\bf 77}, 1413 (1996)

\bibitem{horodecki} M. Horodecki, P. Horodecki and R. Horodecki,
Phys. Lett. A {\bf 223}, 1 (1996)

\end{thebibliography}
\end{document}